\documentclass[10pt]{article}
\usepackage[utf8]{inputenc}
\usepackage[T1]{fontenc}
\usepackage{lmodern}
\usepackage[margin=0.82in]{geometry}
\usepackage{microtype}
\usepackage{graphicx}
\usepackage{booktabs}
\usepackage{tabularx}
\usepackage{array}
\usepackage{float}
\usepackage{pdflscape}
\usepackage[section]{placeins}
\usepackage[font=small,labelfont=bf]{caption}
\usepackage[numbers,sort&compress]{natbib}
\usepackage{xurl}
\usepackage[hidelinks]{hyperref}
\hypersetup{
  pdftitle={SaltyMeta: a curated benchmark and protein language model-informed web tool for salty peptide prediction},
  pdfauthor={Wanchao Chen, Wen Li, Yanan He, Yan Yang},
  pdfsubject={Salty peptide prediction benchmark and web tool},
  pdfkeywords={salty peptide, saltiness-enhancing peptide, protein language model, ESM2, peptide informatics, sodium reduction, benchmark dataset, Streamlit}
}
\newcolumntype{Y}{>{\raggedright\arraybackslash}X}
\title{\textbf{SaltyMeta: a curated benchmark and protein language model-informed web tool for salty peptide prediction}}
\author{Wanchao Chen$^{*}$, Wen Li, Yanan He, and Yan Yang\\
\small Institute of Edible Fungi, Shanghai Academy of Agricultural Sciences\\
\small $^{*}$Corresponding author: \href{mailto:chenwanchao@saas.sh.cn}{chenwanchao@saas.sh.cn}}
\date{}

\begin{document}
\maketitle

\begin{abstract}
Excess sodium intake remains a major public health challenge, while salty and saltiness-enhancing peptides offer a potential route to preserve sensory saltiness in reduced-sodium foods. Machine-learning studies of salty peptides, however, are constrained by small datasets, heterogeneous evidence standards, uncertain negative labels, and sequence similarity leakage. Here we present SaltyMeta, a curated benchmark and web-accessible screening framework for salty or saltiness-enhancing short peptides. The benchmark contains 580 peptides, including 280 positive peptides and 300 negative peptides, all standardized to 2-15 residue one-letter amino-acid sequences. Quality control found no non-standard residues, exact duplicates, or positive-negative overlaps. A similarity-grouped split retained 456 peptides for training and 124 for held-out testing. We evaluated 548 interpretable peptide descriptors, frozen ESM2 embeddings at 8M, 35M, and 150M parameter scales, and descriptor-embedding fusion models under grouped cross-validation. The traditional ExtraTrees baseline selected by training-set grouped cross-validation achieved ROC-AUC=0.693 in cross-validation and ROC-AUC=0.704, PR-AUC=0.702, F1=0.626, and MCC=0.304 on the held-out test set. The best initial protein-language-model fusion was traditional descriptors plus ESM2-8M embeddings, with grouped CV ROC-AUC=0.696 and test ROC-AUC=0.700. Advanced optimization using PCA95 dimensionality reduction and ExtraTrees feature-importance filtering yielded a practical ESM2-8M PCA95 top-300 model with test ROC-AUC=0.715 and PR-AUC=0.703, although repeated-CV gains remained modest. SaltyMeta is therefore positioned as a transparent candidate-prioritization tool rather than a replacement for sensory validation. We provide the benchmark, model package, prediction scripts, GitHub repository, and Streamlit web interface to support reproducible computational screening of food-derived peptides.
\end{abstract}
\noindent\textbf{Keywords:} salty peptide, saltiness-enhancing peptide, protein language model, ESM2, peptide informatics, sodium reduction, benchmark dataset, Streamlit

\section{Introduction}
Reducing sodium intake is an important public health and food-technology goal. The World Health Organization recommends that adults consume less than 2000 mg sodium per day, corresponding to less than 5 g salt per day, and reported that global adult sodium intake remained above this level in 2021 \cite{WHO2026SodiumReduction}. In manufactured and culinary foods, sodium reduction is technically difficult because sodium chloride contributes not only salty taste but also aroma release, texture, preservation, and consumer acceptance. Ingredients that enhance salty perception without proportionally increasing sodium are therefore attractive for reduced-sodium food design.

Taste-active peptides are useful candidates in this setting because they can be generated from food proteins, span diverse sequence chemistries, and may combine taste modulation with nutritional or functional properties. Artificial intelligence platforms for taste peptides now support screening and design across several taste modalities \cite{Yue2025TastepepAI}. Salty peptide prediction is less mature than many other taste-peptide tasks, largely because salty and saltiness-enhancing endpoints are reported with heterogeneous assays and public datasets remain small. In 2026, SaltySought showed that ESM-derived representations combined with machine-learning and deep-learning models can identify and explain salty peptides \cite{Feng2026SaltySought}. SaltyMeta builds on this line of work rather than repeating it: it releases an evidence-stratified salty-peptide benchmark, uses similarity-grouped train/test separation, tests ESM2 scale under small-data constraints, and packages a public web workflow with virtual digestion and peptide-property modules for candidate prioritization.

Short-peptide models are especially prone to leakage. Peptides with high sequence similarity may differ by only one or two residues, so random splits can place nearly identical sequences in both training and test sets. Similarity-aware splitting is now a central issue in biological sequence modeling, including antimicrobial peptide prediction, homology-partitioned benchmarks, and leakage-reduced splitting tools such as GraphPart and DataSAIL \cite{Sidorczuk2022AMPBenchmark,Teufel2023GraphPart,Joeres2025DataSAIL}. For salty peptides, these controls are important because a benchmark with several hundred short sequences can otherwise reward memorization of close sequence variants.

Edible fungi are also an important source domain for SaltyMeta. A recent \textit{Pleurotus eryngii} study identified saltiness-enhancing peptides through integrated screening, sensory evaluation, electronic tongue analysis, molecular docking, and molecular dynamics \cite{Yang2026Pleurotus}. Beyond this report, our research group has published a sequence of edible-fungal taste-peptide studies that identified or screened peptides from \textit{Stropharia rugosoannulata} and \textit{Lentinula edodes} using chromatographic separation, LC-MS/MS, sensory evaluation, electronic-tongue profiling, virtual screening, receptor docking, molecular simulation, and SPR where appropriate \cite{Chen2022Stropharia,Chen2023Lentinula,Li2024StrophariaSPR,Chen2024FoodScience,Chen2025DirectedEnzymolysis}. Although several of these studies were originally framed around umami and taste enhancement, they reported peptides with salty, saltiness-associated, or saltiness-enhancing sensory profiles relevant to the endpoint used here. Eligible peptides from these publications have therefore already been incorporated into the SaltyMeta modelling set, with source provenance retained in Supplementary Table S1. Thus, edible-fungal records in this preprint are part of the literature-curated benchmark rather than a new external validation cohort. Future SaltyMeta work can extend this foundation by expanding the edible-fungal species space, generating independent hydrolysate or peptidomics-derived candidates, ranking them with the frozen model, and validating selected candidates using sensory and physicochemical assays.

Here we present SaltyMeta, a curated benchmark and practical web tool for prioritizing salty or saltiness-enhancing short peptides. The study addresses four questions. First, can the available literature-derived peptide evidence be organized into a transparent benchmark suitable for reproducible modeling? Second, what performance is achieved by conventional interpretable descriptors under similarity-aware evaluation? Third, do frozen protein language model embeddings add useful information in this small-data setting? Fourth, can the resulting model be released as a usable research tool while keeping its limitations visible? The overall workflow is summarized in Figure 1.

\begin{figure}[H]
\centering
\includegraphics[width=0.96\textwidth]{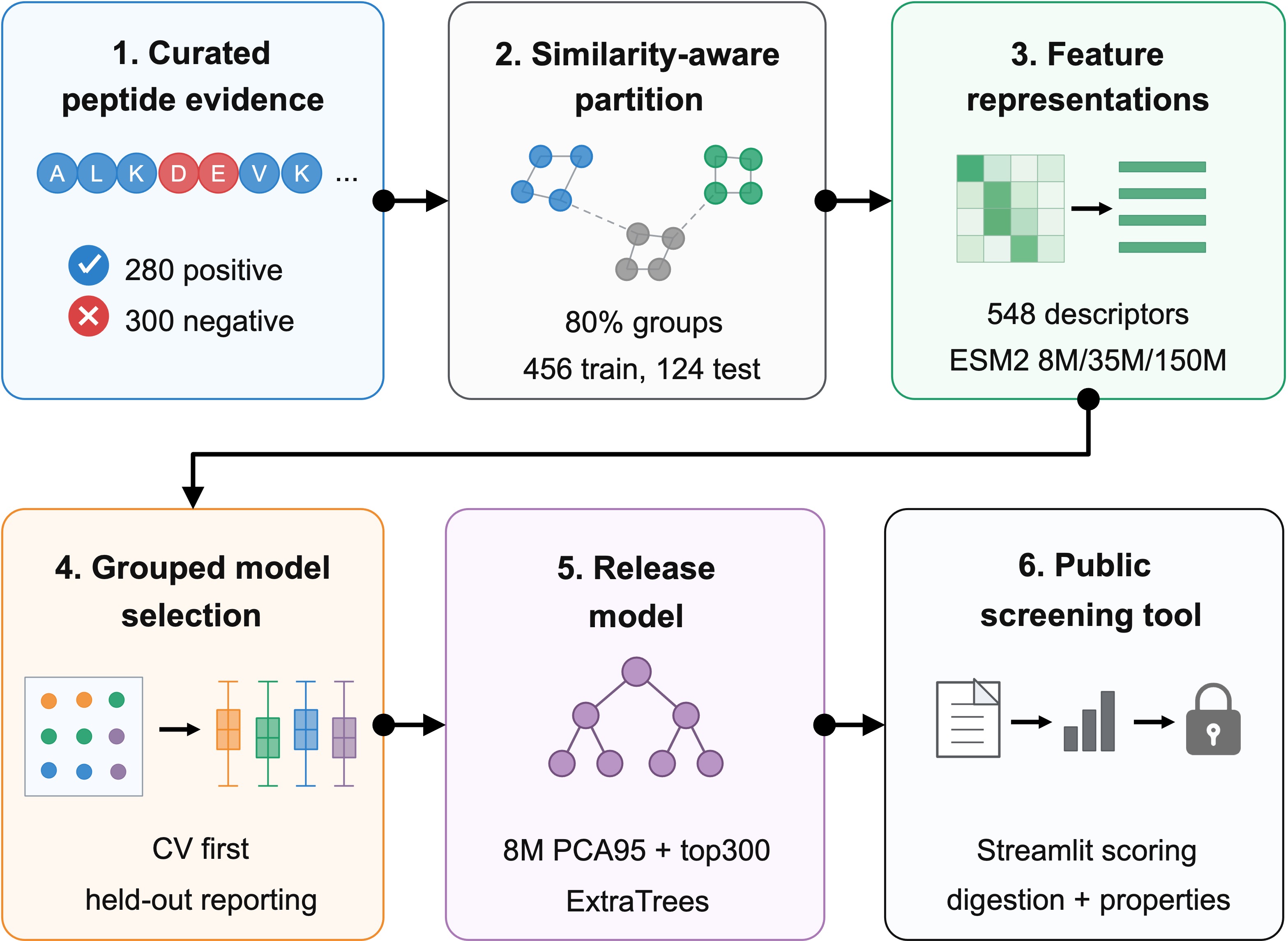}
\caption{SaltyMeta study design, including evidence-retaining curation, similarity-aware splitting, feature extraction, model selection, release-model packaging, and web deployment.}
\label{fig:1}
\end{figure}
\section{Materials and methods}
\subsection{Dataset curation and endpoint definition}
The SaltyMeta benchmark was assembled from the curated positive and negative peptide tables prepared for this project. The positive endpoint was defined as salty or saltiness-enhancing peptide activity. This combined endpoint reflects the reporting structure of the current literature, in which direct salty taste, saltiness enhancement, electronic tongue response, receptor interaction, and database annotation can overlap. The negative class consisted of peptides curated as non-salty or not salty under the source evidence framework.

Sequences were standardized to uppercase one-letter amino-acid codes and restricted to short peptides of 2-15 residues. Quality control checked alphabet validity, peptide length, exact duplicate sequences, positive-negative conflicts, and evidence-field completeness. The final dataset contained 580 unique peptides: 280 positives and 300 negatives. The mean length was 5.86 residues for positives and 5.55 residues for negatives.

Evidence provenance was retained rather than discarded. In the positive table, the QC summary recorded 138 positive entries with direct sensory evidence, 69 entries with literature-derived secondary evidence, and 73 entries with database or compilation annotation. In the negative table, evidence strata included 107 same-study non-salty classifications, 183 database non-salty annotations, 6 human sensory negatives, and 4 electronic-tongue negatives. The expansion audit added 12 direct-sensory positive sequences and 86 sequence-level direct-sensory negative sequences, removed one conflict sequence from both classes, and merged one repeated sequence supported by two concordant sources. These evidence categories were retained for transparency but were not used as model input features in the main classifier.

\subsection{Similarity-aware train/test split}
To reduce leakage from highly similar short peptides, sequences with pairwise similarity at or above 0.80 were grouped before splitting. The grouped split produced a training set of 456 peptides and a held-out test set of 124 peptides. The training set contained 224 positives and 232 negatives; the held-out test set contained 56 positives and 68 negatives. All hyperparameter selection, dimensionality reduction choices, feature filtering, and repeated cross-validation analyses were performed using training-set information only. The held-out test set was reserved for final reporting after model selection.

\begin{table}[tbp]
\centering
\small
\caption{Dataset composition and similarity-aware split.}
\label{tab:1}
\begin{tabular}{lrrr}
\toprule
\textbf{Subset} & \textbf{Non-salty} & \textbf{Salty or saltiness-enhancing} & \textbf{Total} \\
\midrule
Training & 232 & 224 & 456 \\
Held-out test & 68 & 56 & 124 \\
Total & 300 & 280 & 580 \\
\bottomrule
\end{tabular}
\end{table}
The grouped-split strategy follows the general principle that biological sequence benchmarks should avoid placing highly similar sequences across training and evaluation partitions \cite{Sidorczuk2022AMPBenchmark,Teufel2023GraphPart,Joeres2025DataSAIL}. We do not claim that the internal test set is equivalent to a fully external validation set. Instead, the split is used as a leakage-reduced internal evaluation for the current computational release; independent experimental validation is deferred to future work.

\subsection{Traditional peptide descriptors}
For each peptide, we calculated 548 interpretable sequence descriptors. These included peptide length; mass-related descriptors; Kyte-Doolittle hydrophobicity summaries; Hopp-Woods hydrophilicity summaries; approximate net charge at pH 7; charge density; terminal-residue indicators; amino-acid composition; residue counts; grouped residue fractions; dipeptide composition; and composition-transition-distribution descriptors for physicochemical residue groupings. These features were designed to preserve interpretable sequence, composition, charge, size, polarity, and hydrophobicity information.

\subsection{ESM2 protein language model embeddings}
Protein language model representations were extracted from frozen ESM2 models rather than fine-tuned on the salty peptide dataset. ESM-family models learn protein sequence representations from large-scale unsupervised protein data and have shown that biological structure and function information can emerge from protein-language-model scaling \cite{Lin2023ESMFold,Rives2021ProteinLM}. For SaltyMeta, each peptide sequence was embedded using the final hidden layer of the frozen model and mean-pooled across residue positions, excluding special tokens.

Three parameter scales were evaluated: ESM2-8M (esm2\_t6\_8M\_UR50D, 320 dimensions), ESM2-35M (esm2\_t12\_35M\_UR50D, 480 dimensions), and ESM2-150M (esm2\_t30\_150M\_UR50D, 640 dimensions). We intentionally treated model scale as an empirical question. Because the benchmark contains only 580 short peptides, larger embeddings could increase dimensionality and computational cost without improving generalization.

\subsection{Classifiers, metrics, and model selection}
The modeling pipeline evaluated conventional machine-learning classifiers implemented with scikit-learn \cite{Pedregosa2011ScikitLearn}, including logistic regression, support vector machines, random forests \cite{Breiman2001RandomForests}, gradient boosting, k-nearest neighbors, multilayer perceptron models, and ExtraTrees classifiers \cite{Geurts2006ExtraTrees}. For protein-language-model ablation, each ESM2 scale was evaluated as a PLM-only representation and as a fused representation combining traditional descriptors with the corresponding ESM2 embedding.

The primary model-selection criterion was ROC-AUC in five-fold StratifiedGroupKFold cross-validation on the training set. The grouping variable came from the high-similarity clustering described above. We reported held-out ROC-AUC, PR-AUC, accuracy, balanced accuracy, precision, recall/sensitivity, specificity, F1, MCC, and confusion-matrix counts at the default 0.5 threshold. PR-AUC was included because the benchmark is near-balanced but still benefits from precision-recall reporting when the downstream use is candidate prioritization.

\subsection{Advanced optimization and robustness analysis}
After the initial traditional and ESM2 ablation analyses, we evaluated three optimization families under leakage-controlled repeated grouped cross-validation.

First, PCA was applied to high-dimensional PLM blocks as an unsupervised dimensionality-reduction and denoising method \cite{Jolliffe2016PCA}. PCA models were fitted only within the training fold and then applied to the corresponding validation or test data. We tested variance-retention targets including PCA95, which retains 95\% of training-fold variance.

Second, feature-importance filtering was evaluated after descriptor-embedding fusion. ExtraTrees feature importances were estimated from training-fold data, and the top-k features were retained before classifier fitting. This embedded-filtering strategy follows the general feature-selection logic used in high-dimensional supervised learning \cite{Guyon2002FeatureSelection}.

Third, feature-space augmentation was tested with same-class interpolation and mild Gaussian jitter applied only to training folds. Because the class distribution was already near-balanced, augmentation was not presumed to be useful. Synthetic oversampling methods such as SMOTE can be valuable but require caution when synthetic points may not correspond to plausible real biological samples \cite{Chawla2002SMOTE}. Augmentation was therefore treated as exploratory evidence unless it improved repeated grouped cross-validation in a stable way.

The most relevant candidates were then rechecked with 10-seed repeated StratifiedGroupKFold cross-validation. This repeated-CV estimate was used as the main robustness criterion for choosing the practical optimized model. The held-out test set remained a secondary final-reporting estimate.

\subsection{Compact hybrid sensitivity analysis}
We performed a targeted compact-hybrid sensitivity analysis to test whether a lower-dimensional fusion could provide similar or better ranking performance. Traditional descriptors were ranked by ExtraTrees importance within each training fold, the top-k traditional features were retained, and these selected descriptors were concatenated with ESM2-8M PCA95 components. The requested top150 configuration produced 203 total features on the full training set: 150 traditional descriptors and 53 ESM2-8M principal components. Top100, top200, and top250 settings and alternative classifiers were also evaluated to determine whether the 203-feature setting was robust or mainly favorable on the held-out test set.

\subsection{Interpretation}
Permutation importance was computed on the held-out test set for the final traditional ExtraTrees model. Importance results were interpreted as model-level signals rather than direct biochemical mechanisms. Charge distribution, hydrophobicity, polarity, and terminal-residue features were discussed only as plausible correlates of saltiness or saltiness enhancement, especially in light of experimental work linking saltiness-enhancing peptides to sensory response, electronic tongue profiles, and candidate salt-taste targets such as TMC4 \cite{Yang2026Pleurotus}.

\subsection{Web tool and reproducibility package}
The practical optimized model was packaged for reproducible prediction and deployed as a free Streamlit web interface for research-use candidate prioritization. The platform accepts single peptide sequences and batch inputs in common text formats, returns a 0-1 model score, and supports virtual enzymatic digestion and peptide property calculation. The score is treated as a model-derived ranking value, not as a calibrated biochemical probability. Public repository and app links are reported in the availability section.

\subsection{Scope of the current release and future validation boundary}
The current evidence ends at a similarity-grouped internal held-out test and repeated grouped cross-validation. This scope is deliberate: the preprint releases the curated benchmark, modeling pipeline, selected release model, and public screening interface, while reserving independent wet-lab or literature-external validation for a later study. Any future edible-fungal validation should use newly generated or newly curated candidate peptides, remove exact duplicates and close sequence neighbors from the present benchmark, and evaluate the frozen SaltyMeta model without reselecting features, thresholds, or hyperparameters.

\section{Results}
\subsection{Benchmark composition and quality control}
The final SaltyMeta benchmark was near-balanced, with 280 positive and 300 negative peptides. All sequences passed the standard amino-acid alphabet check and the 2-15 residue length restriction. No exact duplicates or positive-negative overlaps were detected after quality control. The similar average lengths of positive and negative peptides reduce the risk that a classifier could rely primarily on length. Figure 2 shows class balance and peptide-length distributions, and Figure 3 summarizes residue-frequency differences by label.

\begin{figure}[tbp]
\centering
\includegraphics[width=0.96\textwidth]{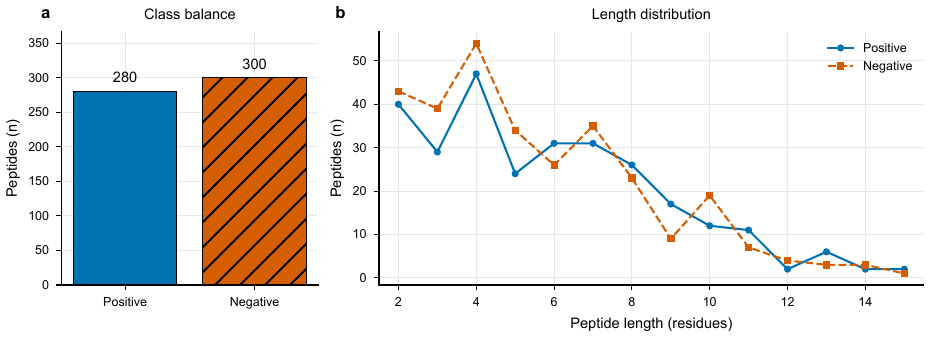}
\caption{Dataset class balance and peptide length distribution in the curated SaltyMeta benchmark.}
\label{fig:2}
\end{figure}
\begin{figure}[tbp]
\centering
\includegraphics[width=0.96\textwidth]{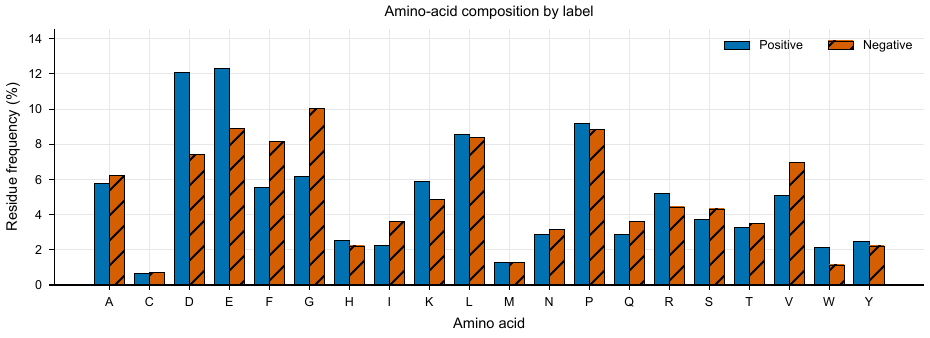}
\caption{Amino-acid frequency by label in the curated benchmark.}
\label{fig:3}
\end{figure}
\begin{table}[tbp]
\centering
\footnotesize
\caption{Evidence-provenance summary retained in the curated benchmark.}
\label{tab:2}
\begin{tabularx}{\linewidth}{>{\raggedright\arraybackslash}p{0.17\textwidth} >{\raggedright\arraybackslash}p{0.27\textwidth} c X}
\toprule
\textbf{Table} & \textbf{Evidence stratum} & \textbf{Count} & \textbf{Interpretation} \\
\midrule
Positive peptides & Direct sensory evidence & 138 & Strongest positive-label evidence in the project QC summary \\
Positive peptides & Literature-derived secondary evidence & 69 & Useful but less direct than sensory assay evidence \\
Positive peptides & Database or compilation annotation & 73 & Retained with provenance for transparency \\
Negative peptides & Same-study non-salty classification & 107 & Stronger negative evidence than database-only annotation \\
Negative peptides & Database non-salty annotation & 183 & Useful but potentially heterogeneous \\
Negative peptides & Human sensory negative & 6 & Direct negative sensory evidence \\
Negative peptides & Electronic-tongue negative & 4 & Instrumental negative evidence \\
\bottomrule
\end{tabularx}
\end{table}
This evidence map is part of the benchmark contribution. It allows future users to construct stricter subsets, such as direct-sensory-only positives or non-database negatives, without rebuilding the entire dataset.

\subsection{Traditional descriptors establish a conservative baseline}
Among traditional-descriptor models, a random forest achieved the highest held-out ROC-AUC of 0.727. The final traditional baseline, however, was not chosen from the test set. Under grouped cross-validation and hyperparameter optimization on the training set, ExtraTrees achieved the best CV ROC-AUC and was selected as the leakage-controlled traditional baseline. Its grouped CV ROC-AUC was 0.693. On the held-out test set, it achieved ROC-AUC=0.704, PR-AUC=0.702, accuracy=0.653, balanced accuracy=0.652, F1=0.626, and MCC=0.304. Figure 4 compares the grouped-CV and held-out ROC-AUC values across the optimized traditional-descriptor models.

\begin{table}[tbp]
\centering
\small
\caption{Traditional ExtraTrees baseline selected by grouped cross-validation.}
\label{tab:3}
\begin{tabularx}{\linewidth}{>{\raggedright\arraybackslash}X r}
\toprule
\textbf{Metric} & \textbf{Value} \\
\midrule
Grouped CV ROC-AUC & 0.693 \\
Held-out ROC-AUC & 0.704 \\
Held-out PR-AUC & 0.702 \\
Accuracy & 0.653 \\
Balanced accuracy & 0.652 \\
Precision & 0.610 \\
Recall/Sensitivity & 0.643 \\
Specificity & 0.662 \\
F1 & 0.626 \\
MCC & 0.304 \\
True negatives/False positives/False negatives/True positives & 45/23/20/36 \\
\bottomrule
\end{tabularx}
\end{table}
This result anchors the comparison. Interpretable peptide descriptors alone support moderate discrimination under similarity-aware evaluation, but the classifier is not strong enough to support stand-alone biological conclusions without experimental validation.

\begin{figure}[tbp]
\centering
\includegraphics[width=0.90\textwidth]{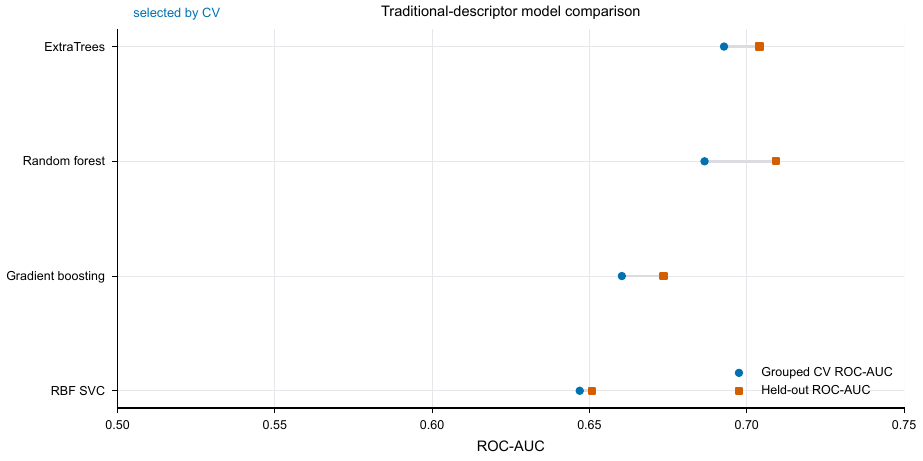}
\caption{ROC-AUC comparison among optimized traditional-descriptor models; ExtraTrees was selected by grouped cross-validation rather than by held-out test performance.}
\label{fig:4}
\end{figure}
\subsection{Frozen ESM2 embeddings add weak complementary information}
The ESM2 ablation showed that protein-language-model scale was not monotonically associated with better salty peptide prediction. The strongest initial fusion by grouped CV was traditional descriptors plus ESM2-8M embeddings with an ExtraTrees classifier. This model reached CV ROC-AUC=0.696 and held-out ROC-AUC=0.700. Relative to the traditional representation in the same PLM ablation table, the CV gain was +0.006, while held-out ROC-AUC decreased slightly from 0.704 to 0.700. Figure 5 summarizes the representation-level comparison.

ESM2-35M fusion reached CV ROC-AUC=0.691 and test ROC-AUC=0.685. ESM2-150M fusion reached CV ROC-AUC=0.689 and test ROC-AUC=0.680. PLM-only models were below the best fusion model, with CV ROC-AUC values from 0.668 to 0.683. These results suggest that frozen ESM2 embeddings contain complementary information, but that this information is weak relative to fold-to-fold uncertainty in the present 580-peptide benchmark. Figure 6 shows the model-scale pattern across the three frozen ESM2 sizes.

\begin{landscape}
\begin{table}[p]
\centering
\scriptsize
\caption{Initial ESM2 feature-set ablation. Fusion models concatenate traditional descriptors with frozen ESM2 embeddings.}
\label{tab:4}
\begin{tabularx}{\linewidth}{>{\raggedright\arraybackslash}X >{\raggedright\arraybackslash}p{2.2cm} c c c c c c}
\toprule
\textbf{Feature set} & \textbf{Best model} & \textbf{Features} & \textbf{Grouped CV ROC-AUC} & \textbf{Test ROC-AUC} & \textbf{Test PR-AUC} & \textbf{F1} & \textbf{MCC} \\
\midrule
ESM2-8M fusion & ExtraTrees & 868 & 0.696 & 0.700 & 0.666 & 0.643 & 0.336 \\
ESM2-35M fusion & ExtraTrees & 1028 & 0.691 & 0.685 & 0.645 & 0.643 & 0.349 \\
Traditional descriptors & ExtraTrees & 548 & 0.690 & 0.704 & 0.700 & 0.644 & 0.324 \\
ESM2-150M fusion & ExtraTrees & 1188 & 0.689 & 0.680 & 0.641 & 0.661 & 0.381 \\
ESM2-8M only & RBF-SVM + PCA & 320 & 0.683 & 0.678 & 0.592 & 0.589 & 0.251 \\
ESM2-35M only & RBF-SVM + PCA & 480 & 0.681 & 0.666 & 0.627 & 0.571 & 0.218 \\
ESM2-150M only & ExtraTrees & 640 & 0.668 & 0.674 & 0.613 & 0.595 & 0.266 \\
\bottomrule
\end{tabularx}
\end{table}
\end{landscape}
\begin{figure}[tbp]
\centering
\includegraphics[width=0.90\textwidth]{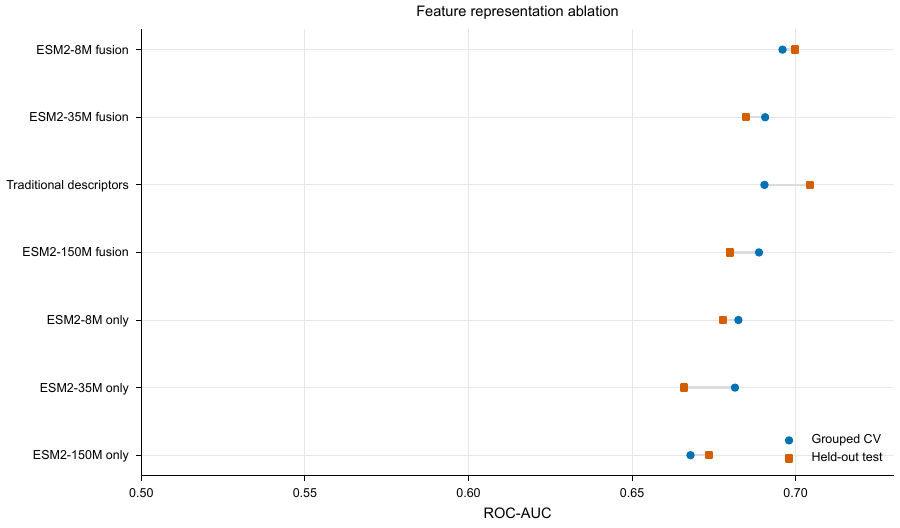}
\caption{Grouped cross-validation and held-out ROC-AUC by feature representation in the initial ESM2 ablation.}
\label{fig:5}
\end{figure}
\begin{figure}[tbp]
\centering
\includegraphics[width=0.90\textwidth]{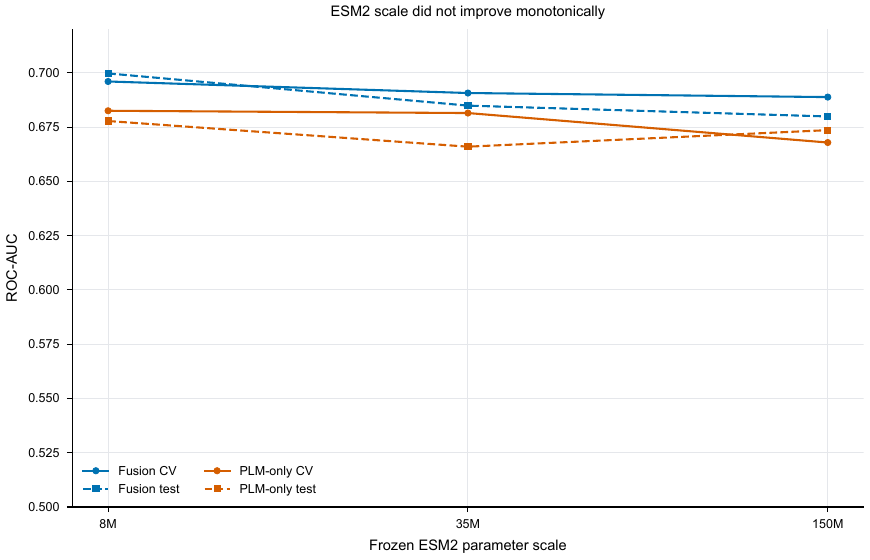}
\caption{Frozen ESM2 parameter-scale ablation across 8M, 35M, and 150M models, showing no monotonic gain from larger embeddings.}
\label{fig:6}
\end{figure}
The result answers the model-scale question pragmatically. For this dataset size and peptide length range, ESM2-8M is the most defensible representation to deploy because it matches or exceeds larger frozen models in grouped CV while reducing embedding cost and feature dimensionality.

\subsection{PCA and feature filtering produce a practical optimized model}
Because raw descriptor-embedding fusion produced only modest improvements, we tested whether PCA denoising and feature-importance filtering could improve robustness. The strict 10-seed repeated grouped-CV leader was ESM2-150M PCA95 fusion, with CV ROC-AUC=0.698$\pm$0.044 and test ROC-AUC=0.714. The practical ESM2-8M PCA95 top-300 model achieved nearly identical repeated CV performance, with CV ROC-AUC=0.697$\pm$0.043, and slightly higher held-out test performance, with ROC-AUC=0.715 and PR-AUC=0.703. Figure 7 shows the repeated-CV/test contrast that drove this conservative selection.

The practical model used traditional descriptors plus ESM2-8M final-layer mean-pooled embeddings reduced by PCA to 95\% training variance. On the full training set, the ESM2-8M PCA95 block contained 53 components. ExtraTrees feature-importance filtering retained the top 300 fused features. The packaged estimator was an ExtraTreesClassifier with 700 trees, balanced class weights, and square-root feature sampling.

\begin{landscape}
\begin{table}[p]
\centering
\scriptsize
\caption{Advanced optimization scenarios. The held-out test column reports ROC-AUC / PR-AUC.}
\label{tab:5}
\begin{tabularx}{\linewidth}{>{\raggedright\arraybackslash}p{4.0cm} >{\raggedright\arraybackslash}X c c >{\raggedright\arraybackslash}p{4.3cm}}
\toprule
\textbf{Scenario} & \textbf{Feature operation} & \textbf{10-seed CV ROC-AUC} & \textbf{Test ROC-AUC / PR-AUC} & \textbf{Decision} \\
\midrule
8M raw fusion reference & 548 traditional + 320 raw 8M features & 0.696$\pm$0.044 & 0.700 / 0.666 & Reference \\
150M PCA95 fusion & 548 traditional + 65 150M PCs & 0.698$\pm$0.044 & 0.714 / 0.699 & Strict CV leader; costlier \\
8M PCA95 + top300 & 548 traditional + 53 8M PCs; top300 retained & 0.697$\pm$0.043 & 0.715 / 0.703 & Selected practical optimized model \\
8M PCA95 + Gaussian jitter & 548 traditional + 53 8M PCs; fold-only jitter & 0.697$\pm$0.043 & 0.717 / 0.707 & Exploratory only \\
8M raw + top300 & 548 traditional + 320 raw 8M features; top300 retained & 0.695$\pm$0.044 & 0.711 / 0.675 & Useful but not selected \\
Traditional top150 & 548 traditional features; top150 retained & 0.674$\pm$0.043 & 0.744 / 0.752 & Not selected because CV was weak \\
\bottomrule
\end{tabularx}
\end{table}
\end{landscape}
\begin{figure}[tbp]
\centering
\includegraphics[width=0.90\textwidth]{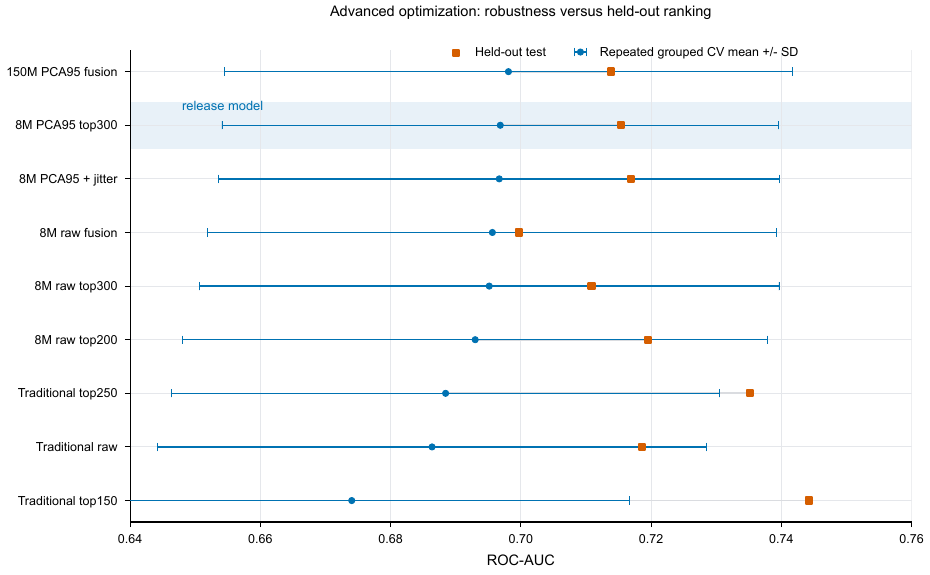}
\caption{Advanced optimization comparison showing repeated grouped-CV ROC-AUC (mean $\pm$ SD) and held-out ROC-AUC.}
\label{fig:7}
\end{figure}
This table illustrates the main interpretive constraint. Several candidates performed well on the held-out test set, including the traditional top150 model with test ROC-AUC=0.744. Those models were not selected when their repeated grouped-CV performance was weaker. Conversely, feature-space Gaussian jitter slightly improved held-out metrics but did not provide a clear repeated-CV advantage over the non-augmented 8M PCA95 top-300 model. We therefore retained jitter as exploratory and selected the non-augmented ESM2-8M PCA95 top-300 ExtraTrees model as the practical release model.

\subsection{Compact hybrid sensitivity supports the conservative selection}
The requested compact hybrid analysis tested whether a simpler fusion of selected traditional descriptors and ESM2-8M PCA95 components could improve ranking. The top150 configuration produced 203 features and achieved held-out ROC-AUC=0.730 and PR-AUC=0.717. However, its 10-seed grouped CV ROC-AUC was 0.690$\pm$0.044, below the practical ESM2-8M PCA95 top-300 model. The best compact-hybrid variant by repeated CV was traditional top250 plus ESM2-8M PCA95 ExtraTrees, with 303 features, CV ROC-AUC=0.698$\pm$0.045, and held-out ROC-AUC=0.711.

These findings support two conclusions. First, compact descriptor-PLM fusion is feasible and may be useful when deployment cost or interpretability constraints matter. Second, the 203-feature top150 model should remain a secondary sensitivity result, not the primary SaltyMeta release model, because test-set improvement did not correspond to stronger repeated grouped-CV evidence.

\subsection{Interpretable descriptors point to charge, polarity, and composition signals}
Permutation importance for the traditional ExtraTrees model highlighted charge-distribution CTD descriptors, FP dipeptide frequency, small-residue fraction, hydrophobicity minimum, size-distribution descriptors, polarity composition, and polar-residue fraction. The strongest feature was ctd\_charge\_distribution\_g1\_p25, followed by ctd\_charge\_distribution\_g1\_p0, dpc\_FP, ctd\_charge\_distribution\_g1\_p100, and group\_frac\_small. Figure 8 displays the ranked importance pattern used for this interpretation.

\begin{table}[tbp]
\centering
\small
\caption{Top held-out permutation-importance features from the traditional ExtraTrees model.}
\label{tab:6}
\begin{tabularx}{\linewidth}{>{\raggedright\arraybackslash}X rr}
\toprule
\textbf{Feature} & \textbf{Importance mean} & \textbf{SD} \\
\midrule
\texttt{ctd\_charge\_distribution\_g1\_p25} & 0.004674 & 0.003332 \\
\texttt{ctd\_charge\_distribution\_g1\_p0} & 0.003834 & 0.003179 \\
\texttt{dpc\_FP} & 0.003184 & 0.002283 \\
\texttt{ctd\_charge\_distribution\_g1\_p100} & 0.002987 & 0.002707 \\
\texttt{group\_frac\_small} & 0.002606 & 0.001770 \\
\texttt{ctd\_charge\_distribution\_g1\_p50} & 0.002600 & 0.001630 \\
\texttt{ctd\_charge\_distribution\_g3\_p75} & 0.002377 & 0.001697 \\
\texttt{kd\_hydrophobicity\_min} & 0.002357 & 0.001443 \\
\texttt{ctd\_size\_distribution\_g1\_p0} & 0.002186 & 0.000696 \\
\texttt{ctd\_polarity\_composition\_g3} & 0.002094 & 0.002095 \\
\bottomrule
\end{tabularx}
\end{table}
\begin{figure}[tbp]
\centering
\includegraphics[width=0.90\textwidth]{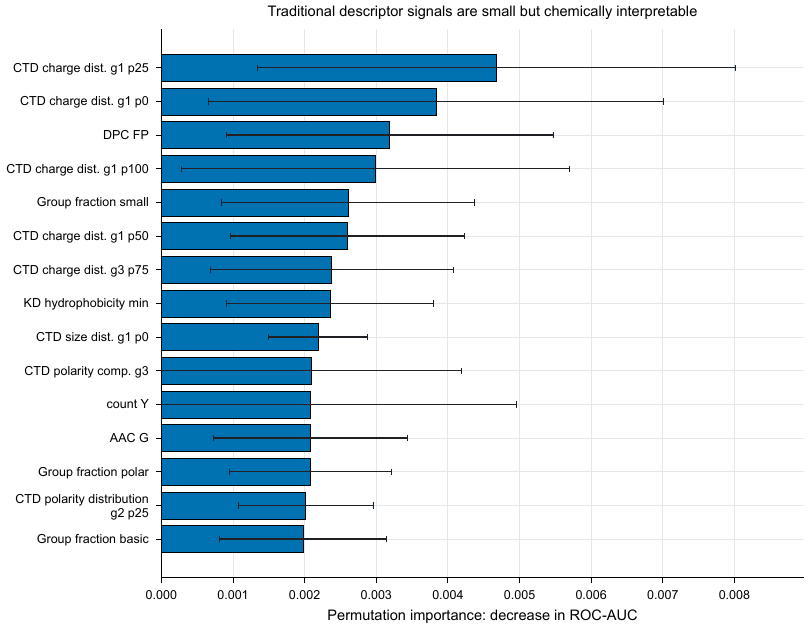}
\caption{Top permutation-importance features from the traditional ExtraTrees model.}
\label{fig:8}
\end{figure}
The prominence of charge-distribution features is chemically plausible, because saltiness perception and saltiness enhancement are expected to depend partly on electrostatic interactions, solubility, and peptide-receptor or peptide-channel contacts. Nevertheless, the importance values are small and model-dependent. They should be interpreted as prioritization clues for follow-up experiments, not as proof of a specific mechanism.

\subsection{SaltyMeta web interface supports candidate prioritization}
The SaltyMeta prediction platform was packaged as a Streamlit web application for research use. The interface supports single peptide prediction, batch peptide scoring, virtual enzymatic digestion, and peptide property calculation. Batch input supports common text formats including CSV, TXT, and FASTA, with the public demo configured for up to 200 sequences per run. The virtual digestion module includes 78 cleavage rules, including food and industrial proteases, and the property toolkit reports values such as molecular weight, pI, pH-specific net charge, GRAVY, residue composition, aromaticity, aliphatic index, extinction estimates, and simple solubility guidance.

The user-facing output is a 0-1 salty prediction score. This score should be used for ranking and triage. It should not be interpreted as a calibrated probability of salty perception, because probability calibration, threshold optimization, and external sensory validation remain future work.

\section{Discussion}
SaltyMeta contributes a reproducible resource and screening workflow for salty peptide discovery. Its main contribution is not a large numerical gain over prior methods. The value of the release lies in evidence-retaining curation, leakage-conscious evaluation, transparent protein-language-model ablation, conservative model selection, and a public tool that researchers can use for candidate triage.

The benchmark addresses a practical problem in taste-peptide informatics. Salty and saltiness-enhancing peptides are often reported through different experimental or database channels, and negative labels can be especially heterogeneous. By retaining evidence strata, the dataset allows users to distinguish direct sensory positives from secondary or database-derived annotations and to recognize that many negatives are not equally strong. This design also makes future sensitivity analyses possible. For example, later versions can train or test on direct-sensory-only subsets once enough labels are available.

The modeling results support restraint in interpreting protein language models on small peptide datasets. ESM-family representations are powerful and have transformed many protein sequence tasks \cite{Lin2023ESMFold,Rives2021ProteinLM}, but the SaltyMeta ablation showed no monotonic advantage from larger frozen ESM2 models. The 8M fusion model was the best initial PLM-assisted model by grouped CV, while 35M and 150M fusion models did not improve held-out performance. After PCA95 dimensionality reduction, the 150M representation became the strict repeated-CV leader by a small margin, but the 8M PCA95 top-300 model offered similar repeated CV, slightly better held-out PR-AUC, and lower embedding cost. For a public pre-screening tool, this practical tradeoff matters.

This conclusion differs from a simplistic "larger model is better" narrative. For short peptides, large frozen embeddings may add many dimensions relative to the number of labeled examples. Without enough independent data, high-dimensional representations can increase variance and create selection instability. The SaltyMeta results suggest a more conservative recipe: use frozen PLM embeddings as complementary descriptors, reduce redundant embedding dimensions, filter fused features within training folds, and select models by grouped resampling rather than by held-out test performance alone.

Compared with SaltySought and broader taste-peptide AI resources \cite{Yue2025TastepepAI,Feng2026SaltySought}, SaltyMeta is best viewed as a complementary release rather than a competing performance claim. Its emphasis is the evidence-stratified benchmark, leakage-conscious evaluation, public candidate-prioritization workflow, and explicit statement of validation boundaries. Cross-paper performance comparisons should be made cautiously because datasets, negative-class definitions, redundancy controls, and validation splits differ. The AUC values near 0.70 reported here may appear modest, but they are more credible for a small heterogeneous literature-curated benchmark than an over-optimized internal score unsupported by independent validation.

The practical optimized model is best viewed as a ranking engine. In a discovery campaign, researchers can generate candidate peptides from hydrolysates or protein databases, score them with SaltyMeta, remove sequences outside the training length range, avoid redundant candidates, and prioritize a feasible synthesis panel. The model can reduce the search space, but it cannot determine sensory quality by itself. Taste is context-dependent: saltiness enhancement may vary with NaCl concentration, pH, matrix composition, peptide purity, bitterness, umami or kokumi side effects, and panel design.

Several limitations should guide use of the current release. First, the positive label combines salty and saltiness-enhancing peptides. These endpoints are related but not identical, and separating them will require larger phenotype-specific datasets. Second, the negative class includes database non-salty annotations that may contain untested or context-dependent activities. Third, the held-out test set is internal, even though similarity grouped; independent external validation remains necessary. Fourth, the default 0.5 threshold is not optimized for a particular experimental budget. Future users should choose thresholds according to synthesis capacity and false-positive tolerance, and should report all tested candidates to avoid success-only reporting. Fifth, the current model does not incorporate peptide concentration, food matrix, assay temperature, or NaCl background, all of which can influence perceived saltiness.

Future edible-fungal validation is better handled as a separate, model-locked study. New mushroom protein hydrolysates can be generated enzymatically, peptides can be identified by LC-MS/MS, and candidate sequences can be filtered to the SaltyMeta training range before model scoring. High-scoring candidates should then be tested by synthesis, HPLC purity confirmation, mass confirmation, electronic tongue analysis, and sensory evaluation under approved protocols. Docking or molecular dynamics against salt-taste targets may help prioritize mechanistic hypotheses but should remain supportive evidence.

\section{Conclusions}
SaltyMeta is a curated benchmark and public screening framework for salty or saltiness-enhancing short peptides. In a 580-peptide dataset evaluated with similarity-aware splits, traditional interpretable descriptors and frozen ESM2 embeddings achieved moderate predictive performance. The strongest practical release model combines traditional descriptors with ESM2-8M embeddings, PCA95 dimensionality reduction, top-300 feature filtering, and an ExtraTrees classifier. It reached held-out ROC-AUC=0.715 and PR-AUC=0.703, while repeated grouped-CV improvements over raw fusion remained small. The model is therefore appropriate for candidate ranking and experimental prioritization, not for definitive sensory classification. The released benchmark, prediction package, and web interface provide a reproducible foundation for computational salty-peptide screening.

\section*{Data and code availability}
Supplementary Tables S1 and S2 provide the curated positive and negative peptide records, evidence annotations, data dictionary, expansion audit, and QC summary. The public SaltyMeta prediction platform is available at \url{https://saltymeta-peptide-predictor.streamlit.app/}. The source code and deployed model package for the web platform are available from the public GitHub repository \url{https://github.com/chenwanchao/saltymeta-prediction-platform}, including the model card and the packaged ExtraTrees model used by the application. The repository commit used for the current web-platform release is 100328e.

\section*{Ethics statement}
This computational study used literature-derived peptide annotations and did not involve new human participants, animal experiments, or clinical data. Future sensory validation of synthesized peptides should be conducted only after the required institutional review, informed consent, food-safety assessment, and allergen or toxicity screening are completed.

\section*{Author contributions}
Wanchao Chen: writing - original draft, conceptualization, formal analysis, and visualization. Wen Li: data curation and methodology; Yanan He: data curation and formal analysis; Yan Yang: writing - review and editing, supervision, and project administration.

\section*{Competing interests}
The authors declare no competing interests.

\section*{AI-use statement}
The manuscript was prepared with assistance from an AI writing tool using author-provided manuscript materials and checked bibliographic metadata. The authors reviewed and remain responsible for the final content, claims, analyses, and references.

\section*{Acknowledgements}
The authors thank the maintainers of the open-source scientific Python ecosystem and the developers of ESM and Streamlit for tools used in this work.

\section*{Supplementary materials}
Supplementary Table S1 contains the curated positive salty or saltiness-enhancing peptide entries. Supplementary Table S2 contains the curated negative peptide entries. The supplementary workbook also includes a README, expansion audit, data dictionary, and QC summary.

\FloatBarrier

\end{document}